\documentclass[sigconf]{acmart}

\usepackage{booktabs} 
\usepackage{amsmath}
\usepackage{amssymb}

\newcommand{\Unif}{\mathrm{Unif}}
\newcommand{\Beta}{\mathrm{Beta}}
\newcommand{\Bern}{\mathrm{Bern}}
\newcommand{\ptrue}{p^{\mathrm{true}}}

\setcopyright{rightsretained}

\acmDOI{10.475/123_4}

\acmISBN{123-4567-24-567/08/06}

\acmConference[KDD 2018]{KDD Conference}{August 2018}{London, England}
\acmYear{2018}
\copyrightyear{2018}

\begin{document}
\title{A Cautionary Tail}
\subtitle{A Framework and Case Study for Testing Predictive Model Validity}

\author{Peter C. Casey}
\orcid{}
\affiliation{%
  \institution{The Lab @ DC\\ Government of the\\ District of Columbia}
  \streetaddress{1350 Pennsylvania Avenue NW}
  \city{Washington}
  \state{D.C.}
  \postcode{20004}
}
\email{peter.casey@dc.gov}

\author{Kevin H. Wilson}
\affiliation{%
  \institution{The Lab @ DC\\ Government of the\\ District of Columbia}
  \streetaddress{1350 Pennsylvania Avenue NW}
  \city{Washington}
  \state{D.C.}
  \postcode{20004}
}
\email{kevin.wilson@dc.gov}

\author{David Yokum}
\affiliation{%
  \institution{The Lab @ DC\\ Government of the\\ District of Columbia}
  \streetaddress{1350 Pennsylvania Avenue NW}
  \city{Washington}
  \country{D.C.}}
\email{david.yokum@dc.gov}

\renewcommand{\shortauthors}{P. Casey, et al.}

\begin{abstract}
Data scientists frequently train predictive models on administrative data. However, the process that generates this data can bias predictive models, making it important to test models against their intended use. We provide a field assessment framework that we use to validate a model predicting rat infestations in Washington, D.C. The model was developed with data from the city's 311 service request system. Although the model performs well against new 311 data, we find that it does not perform well when predicting the outcomes of inspections in our field assessment. We recommend that data scientists expand the use of field assessments to test their models. 
\end{abstract}

%
%


\maketitle

\section{Introduction}
Recently, there has been mounting criticism of the use of predictive analytics in government actions. Much of this is concerned with the notion of fairness \citep{chouldechova2016fairness, hardt2016fairness, kleinberg2016fairness} with a special focus on inequities in data collection \citep{buolamwini2018facialreconition, garvie2016facialrecognition, jackson2017childprotective, larson2016recidivism}. In particular, this research has shown that biased misclassifications can have negative consequences for historically disadvantaged groups when the outcomes being predicted vary with protected classes, like race or gender. This research has typically dealt with criminal justice, education, and employment. In this paper, we deal with similar issues, but in a whole new area of government action: finding rats. 

It may not be possible to extrapolate from even a well-performing model trained on administrative data. The process that generates administrative data is usually designed to fill a specific purpose, different from the purpose for which the data scientist is using it. For this reason, it is important to validate predictive models against their intended use. To do this, we developed a framework for testing predictive models in the field and used it to assess the performance of a model predicting rat infestations in the District of Columbia.

After describing our framework, we present the results of two validations of a model predicting rat infestations. We trained the model with data from D.C.'s 311 service request system, which residents can use to request city services, including rodent abatement. We validated the model against both newly-received 311 requests and the results of a field assessment of 100 inspections. We show that, while our model performs well when predicting the outcomes of new 311 requests, it does not predict the outcomes of inspections well in our field assessment. We encourage applied data scientists to expand the use of field assessments to validate models, especially for government actions.

\section{Background}
Rats are a persistent health risk. Rats carry a wide range of diseases that infect humans \citep{firth2014detection}. Sudden increases in cases of Leptospirosis following recent hurricanes in Puerto Rico and Dominica highlight the impact that rats can have on public health \citep{puertorico2017rats, dominica2015rats}. Rats are especially problematic in urban areas, where large, dense human populations produce waste and contribute to conditions that support rat colonies \citep{himsworth2014mixedmethods}. 

\begin{figure}
\centerline{\includegraphics[width=1.0\linewidth]{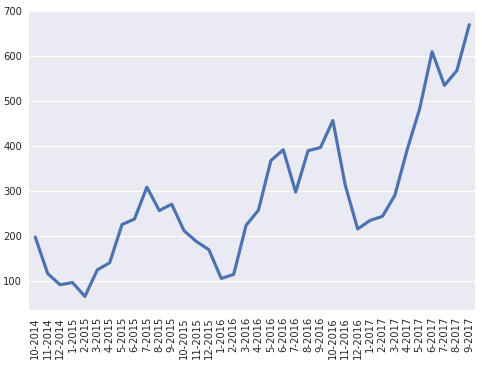}}
  \caption{Number of 311 requests for rodent abatement by month from Oct. 2014 to Sep. 2017. Requests more than doubled from FY2015 to FY2017.}
   \label{fig:rats-by-month}
\end{figure}

In D.C., requests for rodent abatement via the city's 311 system more than doubled from 2,123 in fiscal year 2015 to 5,015 in fiscal year 2017 (Figure \ref{fig:rats-by-month}). When the city receives a 311 request for rodent abatement, the city's rodent control team is deployed to inspect the location.\footnote{Details of rodent control's workflow are based on conversations with the program staff and a ride-along with city pest controllers.} Inspectors generally inspect the entire block where a request was made, searching for burrows that rats use for shelter. If inspectors find rat burrows, they treat the burrows with rodenticide and close the burrows with dirt. Not all requests for rodent abatement lead inspectors to rat burrows. Less than half (46\%) of inspections from August 2015 to August 2017 resulted in rodent control finding rat burrows.

\begin{figure}
\centerline{\includegraphics[width=0.8\linewidth]{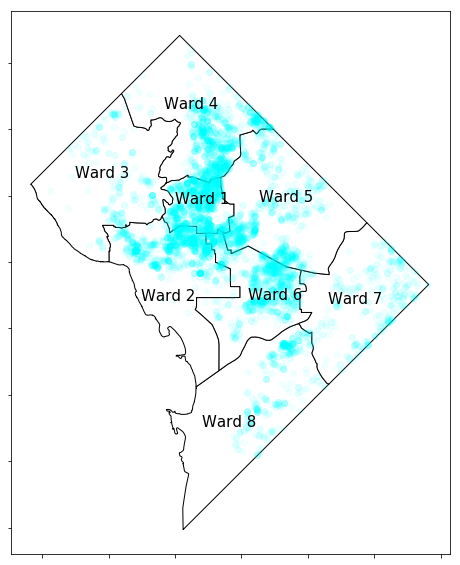}}
  \caption{Requests for rodent abatement in the District of Columbia from FY15 through FY17.}
   \label{fig:rat-map}
\end{figure}

While the 311 system provides rich crowd-sourced data about rodent activity, it is not necessarily an accurate or complete picture of where rodents are in the city. Even when residents witness rodent activity, they may not notify 311. Residents may not know about the city's rodent abatement services, the opportunity cost of contacting 311 may be too high, or they may not believe 311 is an effective way to address the issue. Research has shown that a person's propensity to use 311 at all varies with demographic characteristics that could lead already vulnerable populations to remain underserved \citep{lerman2014staying, levine2014political}. In Washington, D. C., requests for rodent abatement vary geographically (Figure \ref{fig:rat-map}). Requests are most common in the city's densely-populated inner wards, like Ward 1 and Ward 2, and least common in its less dense outer wards, including Wards 3, 7, and 8. 

In order to identify locations where rat infestations may go unreported, we developed a predictive model to extrapolate from incoming 311 data to the environmental factors that contribute to rodent infestations. Our goal was to identify locations in the city where rodent infestations are likely, so that they can be abated regardless of whether or not a 311 request is received. By addressing rodent infestations that go unreported, we hoped to contain and reduce the citywide rodent population and improve the health and safety of District residents.

\section{Field Assessment Framework}
Because we rely on administrative data from D.C.'s 311 system, we were concerned that differences in people's propensity to utilize 311 may bias our model. To ensure that our model performed well in the field, we developed a framework for assessing it. 

First, we selected a number N of locations to inspect based on the model's predictions. We selected N = 100, because this was the number of additional inspections the rodent control team said they could add to their workload.\footnote{The rodent control team committed four code enforcement inspectors to complete the field assessment. Code enforcement inspectors are trained to inspect for rodent activity. However, unlike the inspectors who respond to 311 requests, they are not certified pest controllers. Over the course of the field assessment, rodent control received 311 requests for 25 of the 100 locations selected for the field assessment. Both code enforcement inspectors and certified pest controllers inspected these locations separately. Pest controllers found rat burrows at 16 of the locations, code enforcement found burrows at 14, and the two teams founds burrows at 11 locations in common.} We then selected a range of the distribution of predicted probabilities to test. It was important to have a wide enough range to allow for variation in outcomes across the model's predictions. However, we also wanted to minimize the number of inspections where rodent control did not find rat burrows. For these reasons, we chose to draw our sample from a range of predicted probabilities from 0.5 to 0.9.\footnote{We used a Bayesian test to see if $\mathrm{Pr}(p_1 < p_2) > 0.95$ where the priors on $p_1$ and $p_2$ are both $\Unif[0, 1] = \Beta(1, 1)$. Given these priors, if we see $s$ successes and $N - s$ failures in our data $X_1$, then the posterior $p_1 | X_1 \sim \Beta(s + 1, N - s + 1)$. To test if we had enough power, we modeled  $\ptrue_{1i} \sim \Unif[0.5, 0.6]$ and $\ptrue_{2j} \sim \Unif[0.8, 0.9]$ for $i, j \in [1..N]$ and wrote $X_k = \sum_{i} \Bern(p_{ki})$. We then computed $\mathbb{E}_{X_1, X_2}\left[ \mathrm{Pr}(p_1 | X_1 < p_2 | X_2) > 0.95 \right]$. This value was approximately $0.78$ for $N = 25$.}

We compared the outcomes of the field asessment to predicted probabilities by mapping them on decile plots, where the $x$-axis represents the decile of the predicted probabilities and the $y$-axis represents the percentage of locations in that decile where rat infestations were found. We compared this to a similar mapping of the outcomes of new 311 requests for rodent abatement made over the same period. 

\section{Model Development}
Past models have predicted 311 requests for rodent abatement, relying primarily on features derived from the history of 311 requests in an area \citep{chicago2013rats}. Such an approach risks doubling down on people's uneven propensity to request city services through 311. Instead of predicting the locations of 311 requests, we predict whether or not rat burrows will be found when responding to 311. Rather than use features based on 311 history to predict whether rat burrows will be found, we draw from research in urban rodentology and conversations with the city's rodent control team to develop features based on the environmental factors that contribute to rodent infestations. It was our hope that this approach to modeling rodent infestations would avoid biasing our results towards areas of the city where more city services are requested.

\subsection{Outcome}
Our model predicts the likelihood that inspectors will find rat burrows responding to a 311 request for rodent abatement on a Census block over three months.\footnote{We tested whether the model performed better predicting outcomes over 1, 3, or 6 months during cross-validation. We found that the model always performs better predicting outcomes over 3 months.} We identified the outcomes of inspections by using keywords and phrases to recode inspectors' service notes. If an inspector said they found rat burrows on an inspection, then the inspection was assigned 1, otherwise it was 0. After coding the outcomes of inspections, we  aggregated the results to the Census block. If a rat burrow was found responding to at least one 311 request over three months, the block was assigned 1, otherwise 0. 

We chose Census blocks as our unit of analysis because rats have very limited home ranges, and tend not to cross natural or human-made barriers, such as rivers or roads \citep{traweger2005introducing}. Research has found that even blocks with large infestations may have neighboring blocks with no evidence of rat activity \citep{himsworth2014characteristics}. Since Census blocks tend to correspond to these same natural and human-made barriers, they are a convenient unit of analysis. Census blocks also correspond to the area that rodent control inspects when they respond to 311 requests. 

Restricting our model to predicting the outcomes of 311 requests yields a more balanced sample than predicting the locations of 311 requests, more generally. Requests for rodent abatement only come from about 3\% of the city's Census blocks each month, and less than 30\% of Census blocks from August 2015 to August 2017, the data we use to train the model. In contrast, rodent control finds rat burrows at about 46\% of Census blocks where requests are made.

\subsection{Feature Selection}
We intentionally exclude features derived from 311 requests to avoid biasing our model towards locations where residents may have a higher propensity to request city services. Instead, we limit our features to aspects of the urban environment that contribute to rodent infestation but are unlikely to be correlated with residents' propensity to use 311.\footnote{All features were aggregated to the Census block. For a full list of features, see the model code.} Our decisions about what data to include in the model were guided by research on urban rodentology and conversations with experts, including the city's rodent control team.\footnote{We also consulted extensively with Robert Corrigan, a noted expert on urban rodentology.} 

Like all mammals, rats prefer environments based on the availability of food, water, and shelter \citep{feng2014secret}. Poorly-secured food waste is among the most common sources of food for urban rats \citep{promkerd2008factors, traweger2006habitat}. Denser human populations, the number of residential buildings, and the number of units per building all contribute to trash production and thereby food sources for rodents \citep{himsworth2014mixedmethods}. A mix of residential and commercial zoning can also contribute to rat infestations by increasing trash production, especially when there are nearby restaurants and other food vendors \citep{himsworth2012experiential}.

Norwegian rats, which are the most common in North American cities, like D.C., burrow underground for shelter. Areas with soft earth or deteriorating infrastructure create opportunities for shelter \citep{deMasi2009environmental, himsworth2014mixedmethods}, so we included features measuring impervious surface area and alleyway conditions. Older buildings in poor condition also contribute to rat harborage \citep{himsworth2014mixedmethods}, so we included data on building age and condition. Rats will take shelter in sewers, so we included data on the locations of sewer grates. We also included data about the locations of parks and community gardens where rats may find both food and shelter.

\begin{table}[t]
  \caption{Best Performing Model by Model Type}
  \label{type-perf}
\begin{center}
  \begin{tabular}{l c c c c}
   \toprule
     Model &  Mean & Min & Max & Mean \\
     Type &  P@N &  P@N &  P@N & ROC-AUC \\
    \midrule
    Logistic &  0.56 & 0.50 & 0.64 & 0.60 \\
    Regression & & & & \\
    Gradient &  0.68 & 0.61 & 0.72 & 0.67 \\
    Boosting & & & & \\
    Random &  0.75 & 0.68 & 0.82 & 0.70 \\
    Forests & & & & \\
  \bottomrule
  \end{tabular}
\end{center}
\end{table}

\begin{figure}[t]
\begin{center}
\centerline{\includegraphics[width=1.0\linewidth]{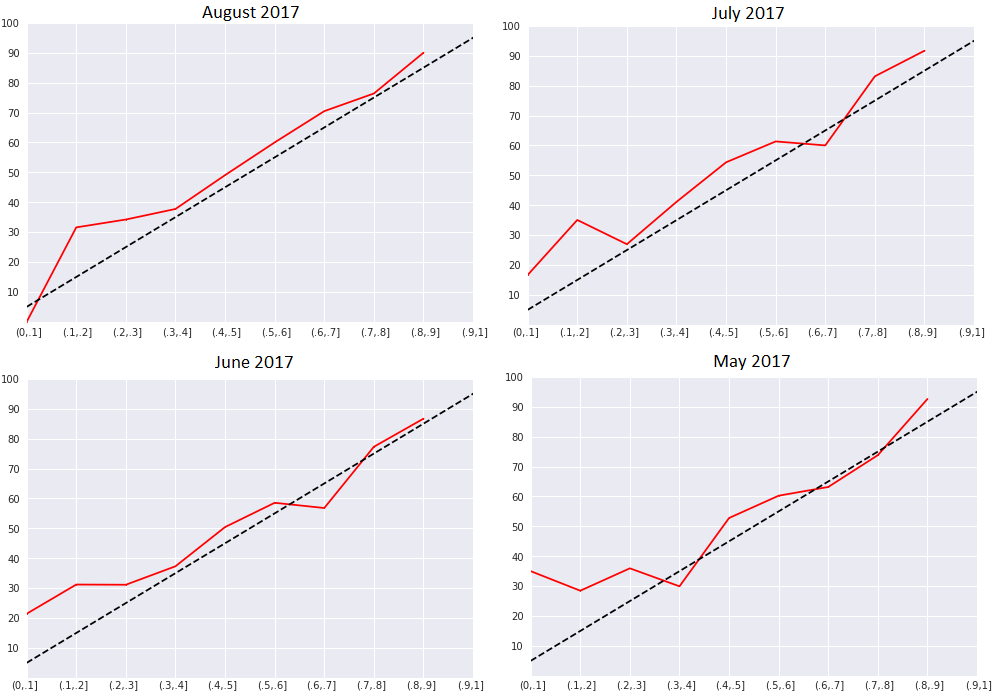}}
   \caption{Decile plots from the last four months of the cross-validation show that the model is well-calibrated to the outcomes of 311 requests. In each subplot, the $x$-axis represents decile bins of the predicted probabilities for each location. The $y$-axis represents the percentage of locations where rat burrows were found. The red line represents the percentage of inspections in each decile bin that rat burrows were found. A perfect relationship between the two axes is represented by the dotted black line. If the model was well-calibrated to the predicted probabilities, we would expect the red line and dotted line to coincide.}
    \label{fig:cv-calibration}
\end{center}
\end{figure}

A wide range of other factors are believed to contribute to rodent infestations. One factor that is commonly cited is construction. We included the locations of recent construction permits in the model. We also included dummies for month and year in order to capture potential seasonal effects. While there is clear seasonal variation in the number of 311 requests for rodent abatement (see Figure \ref{fig:rats-by-month}), research in other cities suggests that rodent populations do not vary significantly seasonally \citep{deMasi2009environmental, himsworth2014characteristics}.

\subsection{Model Selection}

We selected our model type and tuned hyperparameters using temporal cross-validation. Our models were trained with data from August 2015 to August 2017. For each month from August 2016 to August 2017, we trained a model on data from all prior months and tested against the outcomes of inspections over the following three months. We used precision at N (P@N) to evaluate the models, where N = 100. We chose P@N to fit the goal of our model: to maximize the probability that rodent control would find rat infestations at the model's top N targets each month. We chose N = 100 to fit the rodent control team's capacity for additional inspections each month. 

We chose temporal cross-validation for three reasons. First, it replicates our goal of predicting the location of rat infestations in the coming months. Second, we wanted our model to perform well year-round, not only in seasons when requests for rodent abatement are high. Third, since we are always trying to predict the outcomes of inspections on the same set of Census blocks, and the environmental factors in our model features do not change rapidly, autocorrelation is not a particular concern.\footnote{Most of our features are elements of the urban environment that change fairly slowly or not at all. We include proxies for seasons, but the percentage of inspections where rat burrows are found does not vary much seasonally.}

We tested random forest, gradient boosting, and logistic regression models. We found that a random forest classifier performed best (see Table \ref{type-perf}). On average, inspectors found rat burrows at about 74 of the 100 locations identified as most likely to have rat infestations by our chosen model, a more than 60\% improvement over the baseline. The model is also well-calibrated. Figure \ref{fig:cv-calibration} presents decile plots from the last four months of the temporal cross-validation. In each case, the percentage of locations where rat burrows were found fits closely to the predicted probabilities generated from the model. 

\section{Results}
After model selection, we trained the best-performing model on all Census blocks with a 311 request between August 2015 and September 2017 and generated predictions for every Census block for the following three months from October to December. We then selected 100 locations for inspection following the field assessment framework outlined above. From October 10 to November 20, 2017, rodent control completed all 100 inspections and collected data on whether or not they found rat burrows via an online form. We compared the outcomes of these 100 inspections to the model's predictions to assess the its performance. We also compared our model's predictions to the outcomes of 311 requests for rodent abatement over the same period.

\subsection{Field Assessment}

\begin{figure}
\begin{center}
\centerline{\includegraphics[width=0.8\linewidth]{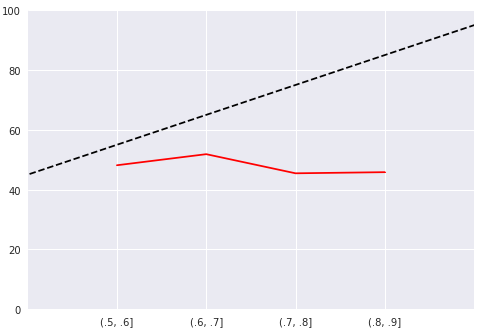}}
  \caption{Decile plot comparing predicted probabilities to field assessment outcomes. See Figure \ref{fig:cv-calibration}. The red line does not coincide with the dotted line, showing that the model is poorly-calibrated to the outcomes of the field assessment.}
    \label{fig:field-validation}
\end{center}
\end{figure}

Figure \ref{fig:field-validation} is a decile plot comparing the outcomes of the 100 inspections in our field assessment to the predicted probabilities generated by the model.\footnote{Running the model 100 times with different random seeds, showed that predicted probabilities were nearly perfectly correlated.} The model's predictions do not fit the outcomes of our field assessment.\footnote{We also not that the AUC was 0.5, which means it was no better than random on the subset of locations inspected. However, because we restricted our field assessment to locations with a high probability of finding rat burrows, we do not believe that AUC is a useful metric.} The percentage of locations where rat burrows were found varies little across the decile bins. Inspectors found rat burrows at 48\% of the locations inspected in the field assessment, ranging from 48\% in decile (0.5, 0.6] to 46\% in decile (0.8, 0.9].

\subsection{Validation against New 311 Requests}

\begin{figure}
\begin{center}
\centerline{\includegraphics[width=0.8\linewidth]{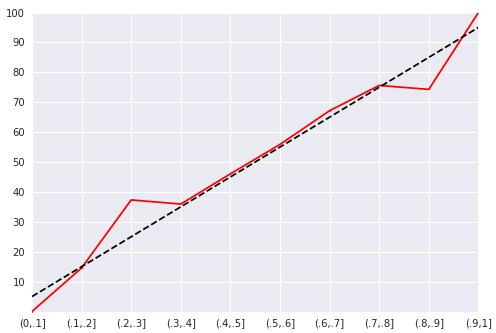}}
  \caption{Decile plot comparing predicted probabilities to 311 request outcomes from Oct. 10 to Nov. 20, 2017. See Figure \ref{fig:cv-calibration}. The red line fits closely to the dotted line, showing that the model is well-calibrated to new 311 requests.}
   \label{fig:311-validation}
\end{center}
\end{figure}

\begin{figure*}[t]
\begin{center}
\centerline{\includegraphics[width=0.8\linewidth]{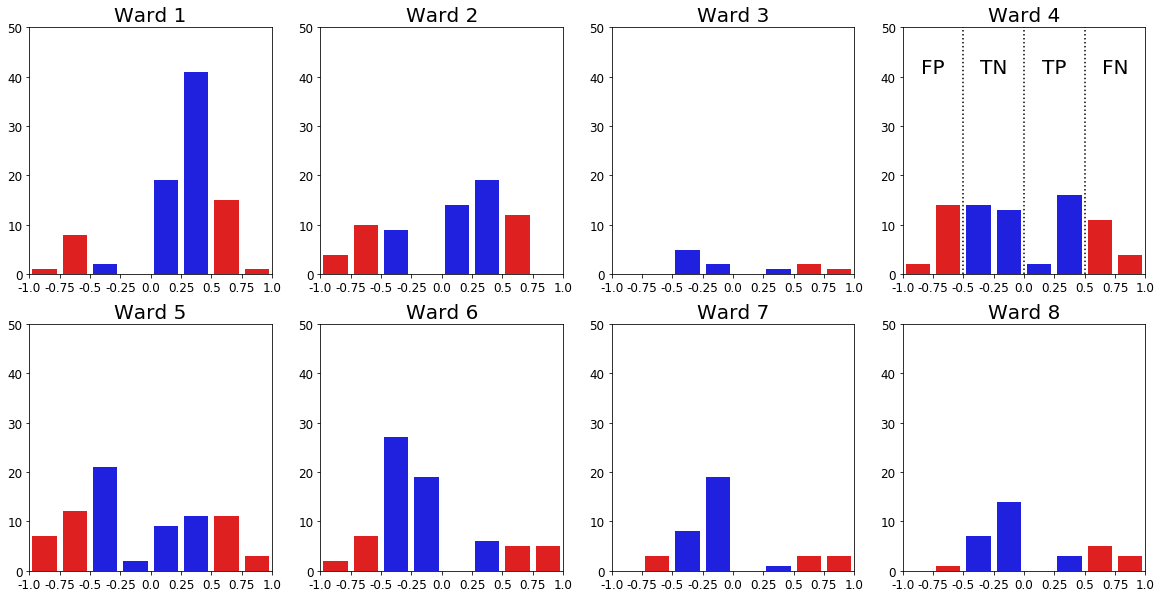}}
  \caption{Distribution of errors by ward in D.C. Errors on the x-axis are the difference between the outcome and the predicted probability for each block in the ward. The y-axis is the number of blocks with an error in that range. Errors are positive where rat burrows were found and negative where rat burrows were not found. Errors below -0.5 indicate False Positives, where as errors above 0.5 indicate False negatives (red). Errors between -0.5 and 0.0 indicated True Negatives, while errors between 0.0 and 0.5 indicated True Positives (blue).}
   \label{fig:errors-by-ward}
\end{center}
\end{figure*}

Figure \ref{fig:311-validation} is a decile plot comparing the model's predictions to the outcomes of 311 requests during the field assessment from October 10 to November 20, 2017. The model was fairly accurate, with AUC at 0.71, and well-calibrated. Pest controllers responding to 311 requests were more likely to find rat burrows in locations with higher predicted probabilities. Rat burrows were found  in 56\% of locations in decile (0.5, 0.6] and 74\% of locations in decile (0.8, 0.9]. 

Figure \ref{fig:errors-by-ward} shows the distribution of model errors by ward in D.C. when compared to the outcomes of 311 requests from October 10 to November 20. The skew of the errors varies by ward, indicating that errors vary geographically. Wards 1 and 2 are densely-populated and located in the downtown center of the city. In both these wards, errors tend to be left-skewed, indicating that the model under-estimated the probability of finding rat burrows in those areas, even though the predicted probabilities in those areas were relatively high. In contrast, errors are right-skewed in Wards 7 and 8, southeast of the Anacostia River, where it is less common to receive reports of rats or find rat burrows. However, when the model is incorrect (i.e., returns a false positive or false negative), errors tend to be slightly left-skewed in those wards. In other words, while the model correctly predicts that rat infestations are uncommon in Wards 7 and 8, where the model is wrong it tends to under-predict the likelihood of finding rat burrows. 

\section{Discussion}
Our results highlight the importance of field testing predictive models. Our model is accurate and well-calibrated when compared to new 311 requests, but the model does not perform well predicting the outcomes of our field assessment. This illustrates the pitfalls of relying on administrative data to both train and validate predictive models. If the data is biased by the process that generates it, then models trained on that data may not be useful in the field. When applied to government actions, biased models can lead to inequitable service delivery. 

We presented a field assessment framework to help address the problems introduced by training models on biased administrative data. To date, there has been little work done to develop similar methods for validating models in the field. We encourage more data scientists to test their models using field assessments, and hope that our methods can serve as a model for such validations. 

\begin{acks}
  The authors would like to thank Mayor Muriel Bowser and her administration, especially City Administrator Rashad Young, Director Jennifer Reed, Director LaQuandra Nesbitt, and interim Chief Technology Officer Barney Krucoff. Also special thanks to Gerard Brown, Sharon Lewis, and the Rodent and Vector Control Division at the District of Columbia's Department of Health for their hard work and expertise, without which this project would not have been possible. Thanks also to Michael Bentivegna, Sam Quinney, Jennifer Doleac, Ryan Moore, Katherine Gan, the Data Team at the Office of the Chief Technology Officer, and The Lab @ DC for their feedback and contributions to the development of this study. Finally, thanks to Robert Corrigan and Donald Green for their expert advice on this project. 
\end{acks}

\bibliographystyle{ACM-Reference-Format}
\bibliography{rats}

\end{document}